\documentclass[aps,prd,twocolumn,preprintnumbers,superscriptaddress,dblfloatfix,nofootinbib]{revtex4-1}
\usepackage[utf8]{inputenc}
\usepackage{lmodern}
\usepackage{amsmath,amssymb}
\usepackage{graphicx}
\usepackage{url}
\usepackage{color}
\usepackage{enumitem}
\usepackage{subfigure}
\usepackage[dvipsnames]{xcolor}
\usepackage[colorlinks=true,breaklinks=true]{hyperref}
\hypersetup{allcolors=[rgb]{0.0 0.0 0.6},linkcolor=[rgb]{0.75 0.05 0.05}}
\usepackage{bm}
\usepackage{epsfig}
\usepackage{amsmath}
\usepackage{amssymb}
\usepackage{slashed}
\usepackage{color}
\usepackage{accents}
\usepackage[dvipsnames]{xcolor}
\usepackage[colorlinks=true,breaklinks=true]{hyperref}
\hypersetup{allcolors=[rgb]{0.0 0.0 0.6},linkcolor=[rgb]{0.75 0.05 0.05}}
\usepackage{orcidlink}

\renewcommand\[{\left[}

\newcommand{\bra}[1]{\langle #1 |}
\newcommand{\ket}[1]{| #1 \rangle}

\newcommand{\TRH}{T_{\rm RH}}
\newcommand{\MPL}{M_P}

\newcommand{\GeV}{\rm GeV}

\newcommand{\exclude}[1]{}

\definecolor{offblue}{RGB}{23,80,153}

\begin{document}

\preprint{IPPP/24/15}

\title{On the Role of Cosmological Gravitational Particle Production in Baryogenesis}
	
\author{Marcos M.~Flores} 
\affiliation{
	Laboratoire de Physique de l'\'{E}cole Normale Sup\'{e}rieure, ENS, Université PSL, CNRS, Sorbonne Universit\'{e}, Universit\'{e} Paris Cit\'{e}, F-75005 Paris, France
}
\author{Yuber F.~Perez-Gonzalez}
\affiliation{Institute for Particle Physics Phenomenology, Durham University, South Road, Durham DH1 3LE, United Kingdom}

\date{\today}
	
\begin{abstract}
    We investigate the generation of the baryon asymmetry within the framework of cosmological gra\-vi\-ta\-tional particle production, employing the Bogoliubov approach. We examine two well-known baryogenesis scenarios, namely baryogenesis in Grand Unified Theories (GUT) and leptogenesis, while considering reheating temperatures sufficiently low for thermal processes to be negligible. Considering $\alpha-$attractor T-models for the inflaton potential, we demonstrate that GUT baryogenesis from scalar decays can be successful across a large range of conformal couplings with gravity, without necessitating substantial levels of CP violation. In the case of leptogenesis, we find that the reheating temperature should be $T_{\rm RH}\lesssim 10^{6}~{\rm GeV}$ for right-handed neutrino masses $M_1 \lesssim 6 \times 10^{12}~{\rm GeV}$ to generate the observed asymmetry.
\end{abstract}

\maketitle
	
\section{Introduction}

The existence of a matter-antimatter asymmetry remains an open question in particle physics and cos\-mo\-lo\-gy. This asymmetry, which cannot be explained by the Standard Model (SM) of particle physics, is one of the largest clues for the necessity for physics beyond our current understanding.

Numerous popular extensions of the SM feature higher dimensional operators which violate baryon or lepton number. Constraints from Planck and the BICEP2/Keck Array indicate that the scale of inflation was less than $1.6\times 10^{16}$ GeV~\cite{Planck:2018jri}, which can be used to establish an upper limit on the reheating temperature of $T\lesssim 10^{9} - 10^{13}$ GeV~\cite{Planck:2018jri, Bezrukov:2007ep, Bezrukov:2011gp}. At these temperatures, the aforementioned baryon-- or lepton-- number violating operators may already be irrelevant. Therefore, the thermal production of any heavy degrees of freedom which violate baryon number and CP would always be exponentially suppressed.

Fortunately, gravity offers the opportunity for saving these scenarios. As already established in the literature, light primordial black holes (PBHs) will evaporate all degrees of freedom which exist~\cite{Hawking:1974rv,Hawking:1974sw}. The heaviest particles will be emitted in the later stages of a black hole's lifetime, as the Hawking temperature exceeds the energy scale of beyond the SM physics. If the particles emitted via Hawking radiation violate baryon or lepton number, as well as the C and CP symmetries, then out of equilibrium decays can lead to generation of a baryon asymmetry. This idea has been explored in many well-motivated models, for example see Refs.~\cite{Toussaint:1978br, Hawking:1980ng, Barrow:1990he, Majumdar:1995yr, Bugaev:2001xr, Baumann:2007yr, Fujita:2014hha, Morrison:2018xla, Ambrosone:2021lsx, Hooper:2020otu, Perez-Gonzalez:2020vnz, Bernal:2022pue, Ghoshal:2023fno}.

While PBHs provide one method for saving high-scale baryogenesis scenarios, this framework is far from minimal. In this paper, we will instead highlight a class of possible baryogenesis scenarios which relay solely on gravity and the cosmological inflationary paradigm. The rapid expansion during the inflationary phase of the early Universe results in the generation of particles through the process known as cosmological gravitational particle production (CGPP). Within the context of particle physics, one can describe the generation of new particles as a result of graviton-mediated inflaton annihilation, which results in the generation of new particles. Alternatively, a non-perturbative approach can be taken which instead relates particle generation to Bogoliubov transformations relating incoming, early time vacuum states to outgoing late time vacuum states. In this context, the generation of particles is analogous to the development of excited states in the non-adiabatic expansion of a one-dimensional square well in quantum mechanics. CGPP can play a crucial role in the evolution of the Universe, and offers a possible origin for cold dark matter, e.g.,~\cite{Chung:1998bt, Kuzmin:1998kk, Chung:1998ua, Chung:1998rq, Chung:2001cb, Ema:2018ucl, Ema:2019yrd, Kaneta:2022gug, Mambrini:2021zpp, Ahmed:2020fhc, Kolb:2023ydq}.

Similar to PBH evaporation, CGPP will produce e\-ve\-ry available degree of freedom, including heavy particles which may be too massive for thermal production. As this is the case, CGPP can lead to the generation of a baryon asymmetry in an identical fashion to PBH evaporation. This possibility has been investigated using the perturbative approach to CGPP, with a particular focus on leptogenesis~\cite{Hashiba:2019mzm, Bernal:2021kaj, Co:2022bgh, Barman:2022qgt, Fujikura:2022udt}. 

In this work we will establish a general framework for calculating the baryon asymmetry generated through CGPP. To do so, we will use the non-perturbative approach to CGPP~\cite{Ema:2018ucl, Ema:2019yrd, Kolb:2023ydq}. Our framework allows for the generation of a baryon asymmetry through the decay of any heavy particle with the necessary properties. However, for concreteness, we will focus on two specific, well-motivated baryogenesis models, namely baryogenesis in Grand Unified Theories (GUT) and leptogenesis. For these two scenarios we have identified the regions of parameter space which can account for the observed matter-antimatter asymmetry.

In Section~\ref{sec:CGPP} we will review the aspects of the non-perturbative CGPP formalism relevant for this work. In Section~\ref{sec:MatterAntimatterAsym} we will discuss the generation of a matter-antimatter asymmetry in the context of CGPP. Lastly, in Section~\ref{sec:Discussion} we will summarize the major results and discuss future directions. Throughout the work, we consider natural units where $\hbar = c = k_{\rm B} = 1$, and define the Planck mass to be $\MPL=1/\sqrt{8 \pi G}$, with $G$ being the gravitational constant.

\section{CGPP Formalism}\label{sec:CGPP}

\subsection{Single-field inflation}

We will focus on single-field, slow-roll inflationary scenario. In particular, we will use a single, real, scalar field which is minimally coupled to gravity through the action,

\begin{equation}
\label{eq:InflatonAction}
S_\varphi = \int d^4 x\sqrt{-g}
\left[
\frac{1}{2}g^{\mu\nu}
\nabla_\mu \varphi
\nabla_\nu \varphi
-V(\varphi)
\right]
.
\end{equation}
For a flat Friedmann-Lemaître-Robertson-Walker (FLRW) space-time,
\begin{equation}
ds^2 = -dt^2 + a^2(t) d^3\mathbf{x}
.
\end{equation}
where $a$ is the scale factor. The Hubble parameter is defined as $H\equiv \dot{a}/a$ where $\cdot \equiv d/dt$. At many points, it is more useful to express our equations in terms of the conformal time $\eta$ which is defined through
\begin{equation}
d\eta = \frac{dt}{a(t)}
.
\end{equation}

The relevant equation of motion, informed by the action, Eq.~\eqref{eq:InflatonAction}, is
\begin{equation}
\ddot{\varphi} + 3H\dot{\varphi} + \frac{dV}{d\varphi} + \Gamma_\varphi\dot{\varphi} = 0,
\end{equation}
where, importantly, we manually have included a decay term to parametrize reheating into radiation. The radiation energy density is determined by the differential equation, 
\begin{equation}
\dot{\rho}_{\rm rad} + 4H\rho_{\rm rad} = (1+w_\varphi)\Gamma_\varphi\rho_\varphi
\end{equation}
where as usual,
\begin{equation}
\rho_\varphi = \frac{1}{2}\dot{\varphi}^2 + V(\varphi),
\end{equation}
and we included the possibility that the inflaton does not behave as matter during reheating with the introduction of the equation-of-state parameter $w_\varphi$.

The evolution of the Hubble parameter is given by
\begin{equation}
3\MPL^2H^2
=
\rho_\varphi + \rho_{\rm rad}
.
\end{equation}
where $\MPL^2 = 1/8\pi G$ is the reduced Planck mass. As is typical, we will define the slow-roll parameter 
\begin{equation}
\varepsilon
\equiv
-\frac{\dot{H}}{H^2}
.
\end{equation}
The end of inflation will be defined as the moment when $\varepsilon = 1$. For the remainder of the paper, we will denote quantities at this time with a subscript $e$. For example, $a_e$, $H_e$ are the scale factor and the Hubble parameter at the end of inflation. Given a scalar potential $V(\varphi)$ and decay constant $\Gamma_\varphi$ one can determine the evolution of the scale factor $a(\eta)$. Following the end of inflation, the inflaton oscillates around the minimum of its potential and decays into radiation during a period known as ``reheating". Once this phase concludes, the Universe is assumed to enter a radiation-dominated era. We define the reheating temperature $T_{\rm RH}$, related to the decay constant $\Gamma_\varphi$, as the radiation temperature when the Universe becomes radiation dominated.
Once the background evolution is determined, $a(\eta)$ will act as an input into the calculations necessary for CGPP.

\subsection{CGPP for spin-0 particles}

Here we will briefly review the physics relevant for our purposes. For a comprehensive review of CGPP, we point the reader to Refs.~\cite{Ema:2018ucl, Kolb:2023ydq}. Consider the action,

\begin{equation}
\label{eq:GenScalarAction}
S_\Phi
=
\int d^4x \sqrt{-g}
\left[
\frac{1}{2}g^{\mu\nu}
\nabla_\mu \Phi \nabla_\nu\Phi
-
\frac{1}{2}m^2\Phi^2
+
\frac{1}{2}\xi\Phi^2 R
\right]
.
\end{equation}
The dimensionless parameter $\xi$ is the coupling between the scalar field $\Phi$ and the Ricci scalar $R$. We will focus on an FLRW space-time and define $\Phi(\eta, \mathbf{x}) = a^{-1}(\eta)\phi(\eta,\mathbf{x})$. Under these assumptions, the action $S_\Phi$ becomes
\begin{align}
S_\Phi^{\rm FLRW}
=
\iint d\eta\ d^3 x
&\left[
\frac{1}{2}(\phi')^2
-
\frac{1}{2}|\nabla\phi|^2
\right.\\[0.25cm]
&- 
\left.
\frac{1}{2} a^2 m_{\rm eff}^2\phi^2
-
\frac{1}{2}
\partial_\eta(aH\phi^2)
\right]
\end{align}
where $' \equiv d/d\eta$ and 
\begin{equation}
m_{\rm eff}
=
m^2 + \left(\frac{1}{6} - \xi \right)R(\eta)
.
\end{equation}
From here we see that if $\xi = 1/6$, the coupling term between the scalar $\Phi$ and the Ricci scalar vanishes. In this circumstance we say that the scalar field is \textit{conformally coupled} to gravity. On the other hand, when $\xi = 0$, the interaction term in Eq.~\eqref{eq:GenScalarAction} vanishes. In this circumstance we say that the scalar field is \textit{minimally coupled} to gravity.

The scalar field equation of motion is given by
\begin{equation}
\phi'' - \nabla^2\phi + a^2 m_{\rm eff}^2\phi = 0
.
\end{equation}
We write the solutions to the above equation as
\begin{equation}
\phi(\eta, \mathbf{x})
=
\int
\frac{d^3 k}{(2\pi)^3}
\left[
a_{\mathbf{k}} \chi_k(\eta) e^{i\mathbf{k}\cdot \mathbf{x}}
+
a_{\mathbf{k}}^\dagger 
\chi_k^*(\eta) e^{-i\mathbf{k}\cdot \mathbf{x}}
\right]. 
\end{equation}
The function, $\chi_k(\eta)$ is the mode function and will play a crucial role in our calculations. Since we are in an FLRW space-time, the mode functions will depend only on the magnitude of the wave number $k = |\mathbf{k}|$.
The mode functions form an orthonormal basis which is illustrated by the fact that they satisfy a Wronskian relation:
\begin{equation}
\label{eq:WronskianCondition}
\chi_k \chi_k'^*
-
\chi_k^* \chi_k'
=
i
.
\end{equation}

The scalar field equation of motion leads to an equation of motion for the mode functions
\begin{equation}
\chi_k''(\eta) + \omega_k^2(\eta)\chi_k(\eta) = 0
\end{equation}
where
\begin{equation}
\omega_k^2(\eta) = k^2 + a^2(\eta)m^2 + \left(\frac{1}{6} - \xi \right)a^2(\eta)R(\eta)
.
\end{equation}

Together with the mode functions, the associated ladder operators $a_{\mathbf{k}}$ and $a_{\mathbf{k}}^\dagger$ define a vacuum state $\ket{0}$ such that $a_{\mathbf{k}}\ket{0} = 0$ for all $\mathbf{k}$. In Minkowski space, it is possible to decompose the mode functions into positive- and negative-frequency components. Subsequently, these mode functions and their associated ladder operators define \textit{unique} vacuum states. In an expanding space-time, this is generally no longer the case. In principle, observers at early times may decompose the mode function in one way, while later observers may select a different decomposition. As a result, the vacuum for one choice of basis may appear as an excited state for another choice.

The two choices of basis functions are related to one another through Bogoliubov transformations of the form
\begin{equation}
\begin{pmatrix}
\tilde{\chi}_k(\eta)\\
\tilde{\chi}_k^*(\eta)
\end{pmatrix}
=
\begin{pmatrix}
\alpha_k & \beta_k\\
\beta_k^* & \alpha_k^*
\end{pmatrix}
\begin{pmatrix}
\chi_k(\eta)\\
\chi_k^*(\eta)
\end{pmatrix}
.
\end{equation}
The ladder operators transform as
\begin{equation}
\begin{pmatrix}
\tilde{a}_{\bf k}\\
\tilde{a}_{-\bf k}^\dagger
\end{pmatrix}
=
\begin{pmatrix}
\alpha_k^* & -\beta_k^*\\
-\beta_k & \alpha_k
\end{pmatrix}
\begin{pmatrix}
a_{\bf k}\\
a_{-\bf k}^\dagger
\end{pmatrix}
.
\end{equation}

Here the complex entries of the matrix above satisfy $|\alpha_k|^2 \pm |\beta_k|^2 = 1$, with $-$ for bosons and $+$ for fermions. In particular, the early time or incoming mode functions can be expressed in terms of the late time or outgoing mode functions through
\begin{equation}
\chi_k^{\rm IN}(\eta)
=
\alpha_k\chi_k^{\rm OUT}(\eta)
+
\beta_k\chi_k^{\rm OUT *}(\eta)
.
\end{equation}
Using Eq.~\eqref{eq:WronskianCondition} we can find the Bogoliubov coefficients which connect the incoming and outgoing mode functions
\begin{equation}
\begin{split}
\alpha_k
&=
i
(
\chi_k^{\rm OUT *}\chi_k'^{\rm IN}
-
\chi_k'^{\rm OUT *}\chi_k^{\rm IN}
),\\[0.25cm]
\beta_k
&=
i
(
\chi_k^{\rm OUT}{}'\chi_k^{\rm IN}
-
\chi_k^{\rm OUT}\chi_k^{\rm IN}{}'
).
\end{split}
\end{equation}
Similarly,
\begin{equation}
a_{\bf k}^{\rm IN}
=
\alpha_k^* a_{\bf k}^{\rm OUT}
+
\beta_k^* a_{-\bf k}^{\rm OUT}{}^\dagger
.
\end{equation}

As is traditionally the case in examinations of inflation, we will associate the initial vacuum state as the Bunch-Davies vacuum~\cite{Bunch:1978yq}, $\ket{0}_{\rm BD}$, such that $a_{\mathbf{k}}^{\rm IN}\ket{0}_{\rm BD} = 0$ for all $\mathbf{k}$. For a system initially in the Bunch-Davies vacuum, the number operator, as specified by the outgoing ladder operators is given by
\begin{equation}
N^{\rm out}
=
\int
\frac{d^3k}{(2\pi)^3}
a_{\mathbf{k}}^{\rm OUT}{}^\dagger
a_{\mathbf{k}}^{\rm OUT}
.
\end{equation}
The number of particles measured by the above operator in the Bunch-Davies vacuum is
\begin{equation}
{}_{\rm BD}
\bra{0}N^{\rm OUT}\ket{0}_{\rm BD}
=
V
\int
\frac{d^3 k}{(2\pi)^3}|\beta_k|^2
\end{equation}
where $V$ is the expected IR divergence $V = (2\pi)^3\delta(\mathbf{k} - \mathbf{k})$. From here, we can deduce the comoving number density of particles as measured in the outgoing basis:
\begin{equation}
a^3 n
=
\int
\frac{d^3k}{(2\pi)^3}
|\beta_k|^2
.
\end{equation}
The above equation can be rewritten as
\begin{equation}
a^3 n 
=
\int
\frac{dk}{k}\ a^3 n_k,
\quad
{\rm with}
\quad
a^3 n_k
=
\frac{k^3}{2\pi^2}
|\beta_k|^2
.
\end{equation}
Here, $a^3n_k$ is the comoving spectral number density and is the key quantity determined within the framework of CGPP. 

In order to find $a^3 n_k$ we will adopt the parametrization of Ref.~\cite{Kofman:1997yn},
\begin{equation}
\label{eq:InModeParamet}
\chi_k^{\rm IN}(\eta)
=
\frac{\tilde{\alpha}_k(\eta)}{\sqrt{2\omega_k(\eta)}} e^{-i\Phi_k(\eta)}
+
\frac{\tilde{\beta}_k(\eta)}{\sqrt{2\omega_k(\eta)}} e^{i\Phi_k(\eta)}
\end{equation}
where $\Phi_k'(\eta) = \omega_k(\eta)$. Inserting the above parametrization into the mode equation of motion leads to a set of three, coupled first-order differential equations
\begin{subequations}\label{eq:ScalarModeEqns}
\begin{align}
\partial_\eta \tilde{\alpha}_k(\eta) 
&=
+\frac{1}{2}A_k^\Phi(\eta)\omega_k(\eta)\tilde{\beta}_k(\eta) e^{+2i\Phi_k}\\[0.25cm]
\partial_\eta \tilde{\beta}_k(\eta) 
&=
+\frac{1}{2}A_k^\Phi(\eta)\omega_k(\eta)\tilde{\alpha}_k(\eta) e^{-2i\Phi_k}\\[0.25cm]
\partial_\eta\Phi_k(\eta) &= \omega_k(\eta)
\end{align}
\end{subequations}
where
\begin{equation}
A_k^\Phi(\eta) 
= 
\frac{\omega_k'(\eta)}{\omega_k^2(\eta)}
\end{equation}
is the adiabaticity in the context of scalar CGPP. In FLRW space-times, the adiabaticity parameter vanishes as $\eta\to\pm \infty$. This fact enables us establish initial conditions for the above system of differential equations (see Refs.~\cite{Ema:2018ucl, Kolb:2023ydq})
\begin{equation}\label{eq:ModeICs}
\begin{split}
\tilde{\alpha}_k(\eta\to -\infty) &= 1,\\
\tilde{\beta}_k(\eta\to -\infty) &= 0,\\
\Phi_k(\eta\to -\infty) &= 0.
\end{split}
\end{equation}
Furthermore, one can establish that the Bogoliubov transformation coefficient needed to determine the comoving number density is simply,~\cite{Ema:2018ucl, Kolb:2023ydq}
\begin{equation}
\beta_k
=
\lim_{\eta\to\infty}
\tilde{\beta}_k(\eta)
.
\end{equation}

\textit{In summary}, in order to determine the comoving number density of particles produced through CGPP one must first determine the background evolution, i.e., $a(\eta)$. After doing so, solving the system of differential equations, Eq.~\eqref{eq:ScalarModeEqns}, subject to the initial conditions, Eq.~\eqref{eq:ModeICs}, will yield $\beta_k
$ after evaluating the solutions obtained at late conformal times. We will offer an illustrative example of this procedure after discussing the formalism required to describe the generation of spin-1/2 particles.

\subsection{CGPP for spin-1/2 particles}

Spiritually, the analysis for spin-1/2 particles is similar to the scalar case. However, some subtitles arise due to the fact that the spin-1/2 fields are represented by spinors. The gravity-based generation of spin-1/2 particles has been examined and reviewed in Refs.~\cite{Kuzmin:1998kk, Chung:2011ck, Boyle:2018rgh, Ema:2019yrd, Kolb:2023ydq}. For completeness, we will summarize the formalism required to examine the implications of spin-1/2 CGPP for baryo/leptogenesis.

The coupling of these particles to gravity is given by the action
\begin{equation}
S_\Psi = \int d^4 x\ \sqrt{-g}
\left[
\frac{i}{4}
\left(
\bar{\Psi}
\gamma^\mu \nabla_\mu \Psi
-
\bar{\Psi}
\overset{\leftarrow}{\nabla}_\mu \gamma^\mu
\Psi
\right)
-
\frac{1}{2}m\bar{\Psi}\Psi
\right]
.
\end{equation}
We will impose $\Psi = \Psi^C\equiv -i\gamma^2\Psi^*$ implying that $\Psi$ is a self-conjugate Majorana spinor. To accommodate particles with half-integer spin within the framework of general relativity, one must use the frame-field formalism. This formalism extends field theory for spinors into general curved space-times. For simplicity, we will point the reader to Refs.~\cite{Parker:2009uva, Freedman:2012zz} for further details on this topic.

For the particular case of an FLRW metric, the equation of motion corresponding to the above action can be found to be
\begin{equation}
\label{eq:DiracEqn}
(i\gamma^{a}\delta_{a}^{\mu} \partial_{\mu} - a(\eta)m)
\left[
a^{3/2}(\eta)\Psi(\eta, \mathbf{x}) 
\right]
.
\end{equation}
Taking $\psi = a^{3/2}\Psi$ recovers the Minkowski space-time Dirac equation with a time-dependent mass $a(\eta)m$. The solution to the FLRW field equation of motion is given by
\begin{multline}
\label{eq:FermionFLRW}
a^{3/2}(\eta)\Psi(\eta, \mathbf{x})
=\\[0.25cm]
\int \frac{d^3k}{(2\pi)^3}
\sum_{\lambda = \pm 1/2}
\left[
a_{\mathbf{k},\lambda} U_{\mathbf{k},\lambda}(\eta, \mathbf{x})
+
a_{\mathbf{k},\lambda}^\dagger V_{\mathbf{k},\lambda}(\eta, \mathbf{x})
\right]
\end{multline}
where the mode functions $U_{\mathbf{k},\lambda}$ are given by
\begin{equation}
U_{\mathbf{k},\lambda}(\eta, \mathbf{x})
=
\begin{pmatrix}
u_{A,k}(\eta)\\
\pm u_{B,k}(\eta)
\end{pmatrix}
\otimes
h_{\mathbf{k}, \pm}\ 
e^{i\mathbf{k}\cdot \mathbf{x}}
\end{equation}
and $V_{\mathbf{k}, \lambda} = -i\gamma^2 U_{\mathbf{k},\lambda}^*$. Both the mode functions and the ladder operators are labeled by comoving wavevector $\mathbf{k}$ and helicity $\lambda$. The two component spinors, $h_{\mathbf{k},\lambda}$ are choosen such that
\begin{equation}
\frac{1}{2}
(
\hat{\mathbf k}\cdot \boldsymbol{\sigma}
)
h_{\mathbf{k},\pm}
=
\pm
\frac{1}{2}h_{\mathbf{k},\pm},
\end{equation}
where $\boldsymbol{\sigma}$ represent the three Pauli matrices, and are normalized such that
\begin{equation}
h_{\mathbf{k},2\lambda}^\dagger
h_{\mathbf{k},2\lambda'}
=
\delta_{\lambda\lambda'}
.
\end{equation}
Using the solution Eq.~\eqref{eq:FermionFLRW} with the equation of motion Eq.~\eqref{eq:DiracEqn} leads to a coupled differential equation for the time-dependent components $u_{A,k}(\eta)$ and $u_{B,k}(\eta)$,
\begin{equation}
i\partial_\eta
\begin{pmatrix}
u_{A,k}\\
u_{B,k}
\end{pmatrix}
=
\begin{pmatrix}
a(\eta) m & k\\
k & -a(\eta) m
\end{pmatrix}
\begin{pmatrix}
u_{A,k}\\
u_{B,k}
\end{pmatrix}
.
\end{equation}
The time-dependent eigenvalues of the matrix above are given by
\begin{equation}
\omega_k(\eta) = \pm\sqrt{k^2 + a^2(\eta)m^2}
.
\end{equation}
In analogy with the scalar case, Eq.~\eqref{eq:InModeParamet}, we can parameterize the incoming spinor components $u_{A,k}$ and $u_{B,k}$ as
\begin{equation}
\begin{pmatrix}
u_{A,k}^{\rm IN}\\
u_{B,k}^{\rm IN}
\end{pmatrix}
=
\mathbf{M}
\begin{pmatrix}
\tilde{\alpha}_k\\
\tilde{\beta}_k
\end{pmatrix}
\end{equation}
with 
\begin{equation}
\mathbf{M}
=
\begin{pmatrix}
\sqrt{\frac{\omega_k + a(\eta)m}{2\omega_k}}
 e^{-i\Phi_k(\eta)}
 & -\sqrt{\frac{\omega_k - a(\eta)m}{2\omega_k}}
 e^{i\Phi_k(\eta)}\\[0.25cm]
\sqrt{\frac{\omega_k - a(\eta)m}{2\omega_k}}
 e^{-i\Phi_k(\eta)} & \sqrt{\frac{\omega_k + a(\eta)m}{2\omega_k}}
 e^{i\Phi_k(\eta)}
\end{pmatrix}
.
\end{equation}
where $\Phi_k(\eta)$ is defined as before. This redefinition leads to a system of three coupled differential equations:
\begin{subequations}\label{eq:SpinHalfModeEqns}
\begin{align}
\partial_\eta \tilde{\alpha}_k(\eta) &= 
-\frac{1}{2}
A_k^\Psi(\eta)
\omega_k(\eta)\tilde{\beta}_k(\eta) e^{+2i\Phi_k}\\[0.25cm]
\partial_\eta \tilde{\beta}_k(\eta) 
&= 
+\frac{1}{2}
A_k^\Psi(\eta)
\omega_k(\eta)\tilde{\alpha}_k(\eta) e^{-2i\Phi_k}\\[0.25cm]
\partial_\eta\Phi_k(\eta) &= \omega_k(\eta)
\end{align}
\end{subequations}
where
\begin{equation}
A_k^\Psi(\eta) 
= 
\frac{a^2 H mk}{\omega_k^3(\eta)}
\end{equation}
is the adiabaticity in the context of spin-1/2 CGPP and is, importantly, different than that derived in the scalar case. Also note that there is a difference in sign between Eq.~\eqref{eq:ScalarModeEqns} and Eq.~\eqref{eq:SpinHalfModeEqns}. The mass dependence of the spin-1/2 adiabaticity demonstrates that only massive spin-1/2 particles can be generated gravitationally. 
\begin{figure*}[t]
    \includegraphics[width=0.9\linewidth]{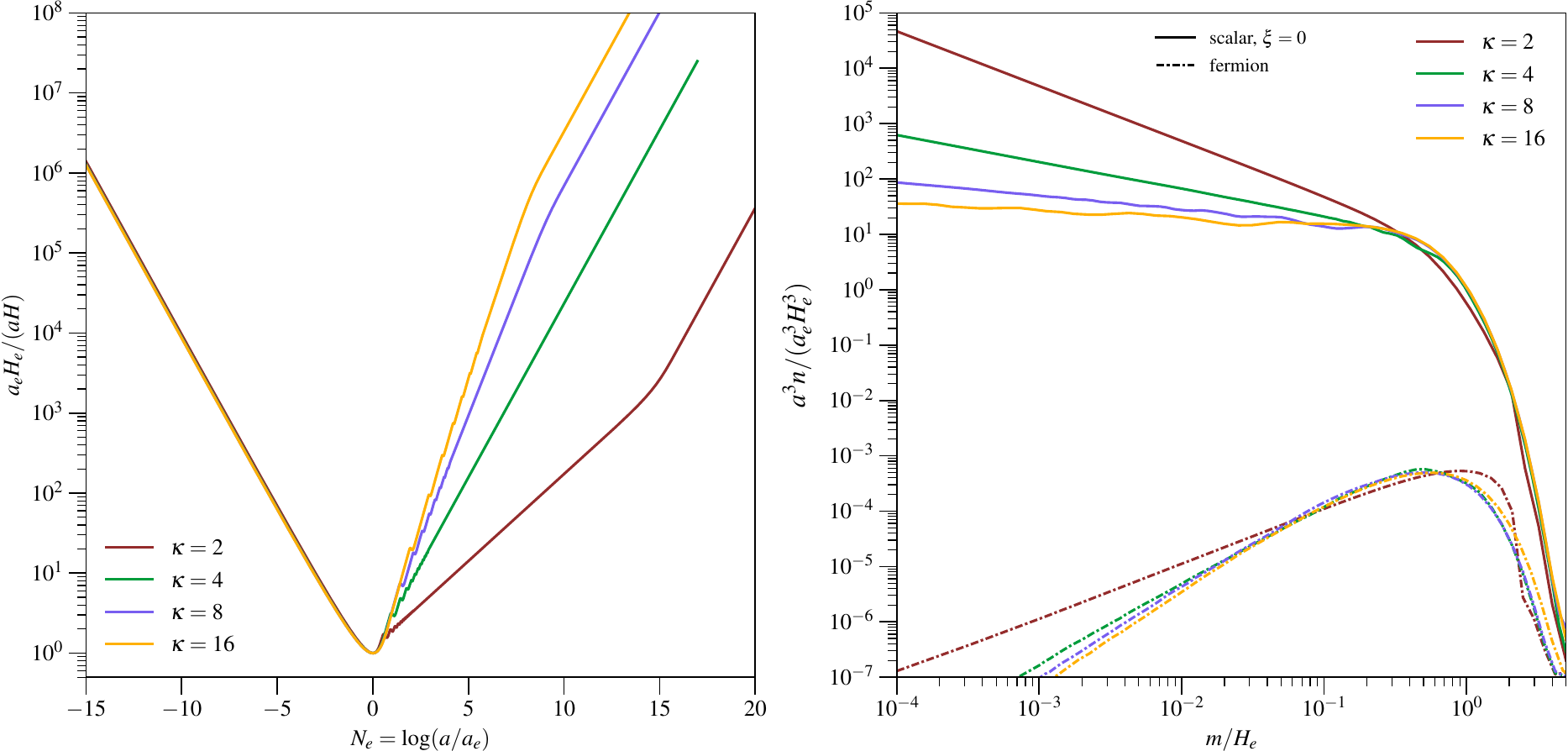}
    \caption{Left. Evolution of the comoving Hubble radius normalized by the value at the end of inflation. Right. Comoving number densities from CGPP for scalars with $\xi=0$ (full) and fermion (dot-dashed), as function of the mass particle normalized to Hubble at end of inflation. We present the predicted abundances for $\kappa=2$ (red), $\kappa=4$ (green), $\kappa=8$ (light purple), and $\kappa=16$ (light orange).}
	\label{fig:HubbleRadius}
\end{figure*}

From here we can see that, procedurally, determining the fermion number density is identical to the scalar case up to small differences in the mode equations and the spin-1/2 adiabaticity. Having reviewed the CGPP formalism, we will now proceed to use these tools for determining the number densities relevant for baryogenesis.

\subsection{Implementation with $\alpha$-attractor T-model}

We will now proceed with an explicit example, and calculate particle number densities using the formalism described above. As discussed previously, the two major steps required are (i) determine the background evolution and (ii) using this solution, solve the mode equations. 

For our particular case, we will consider the $\alpha$-attractor T-model~\cite{Kallosh:2013hoa},
\begin{equation}\label{eq:potential_inflaton}
V(\varphi) = \lambda \MPL^4
\left|
\sqrt{6}\tanh
\left(
\frac{\varphi}{\sqrt{6} \MPL}
\right)
\right|^{\kappa}
.
\end{equation}
A potential of this form appears within the context of no-scale super gravity and is consistent with modern-day observations of the cosmic microwave background~\cite{Planck:2018jri}. 
The parameter $\lambda$ can be related directly to the measured amplitude of the curvature power spectrum through~\cite{Garcia:2020wiy,Co:2022bgh}
\begin{equation}
\lambda = \frac{18\pi^2}{6^{\kappa/2}N_*^2}\ A_{S*}
\end{equation}
where is measured to be $A_{S*} = 2.101\times 10^{-9}$~\cite{Planck:2018jri, Planck:2018vyg} and the number of $e$-folds $N_*$ is taken to be $N_* = 55$ for the Planck pivot scale $k_* = 0.05$ Mpc${}^{-1}$~\cite{Co:2022bgh, Garcia:2020wiy}. Near the origin, the above potential is approximately
\begin{equation}
V(\varphi) 
\approx
\lambda
\frac{\varphi^\kappa}{\MPL^{\kappa - 4}}
.
\end{equation}
Due to this behavior, the oscillation-averaged inflaton energy density can scales as
\begin{equation}
\langle
\rho_\varphi(a)
\rangle_{\rm osc.}
=
\rho_e
\left(
\frac{a_e}{a}
\right)^{\frac{6\kappa}{\kappa + 2}}
.
\end{equation}
The equation-of-state parameter during this oscillatory phase is given by $w_\varphi = (\kappa - 2)/(\kappa + 2)$~\cite{Garcia:2020wiy}. As we will discuss in the next section, changes in the parameter $\kappa$, and therefore the evolution of the reheating period, have drastic implications for the generation of a baryon asymmetry through CGPP.

We present the results of the background solution calculation and the CGPP calculation in Fig.~\ref{fig:HubbleRadius} for various values of $\kappa$. On the left panel of Fig.~\ref{fig:HubbleRadius} we have plotted the normalized comoving Hubble radius. From here, one can see the different equations of states which result from adjusting the value of $\kappa$. All of these phases end in a radiation dominated era as is required for the standard cosmological timeline. The left panel features the comoving number densities for minimally coupled scalars and fermions. The large relative differences between the minimally coupled scalar and the fermionic number densities will play a significant role in our discussion of viable models of baryogenesis. 

\section{Generation of matter-antimatter asymmetry}\label{sec:MatterAntimatterAsym}

\subsection{Asymmetry evolution in $\alpha$-attractor T-models during reheating}

\begin{figure}[!t]
    \centering
    \includegraphics[width=0.9\linewidth]{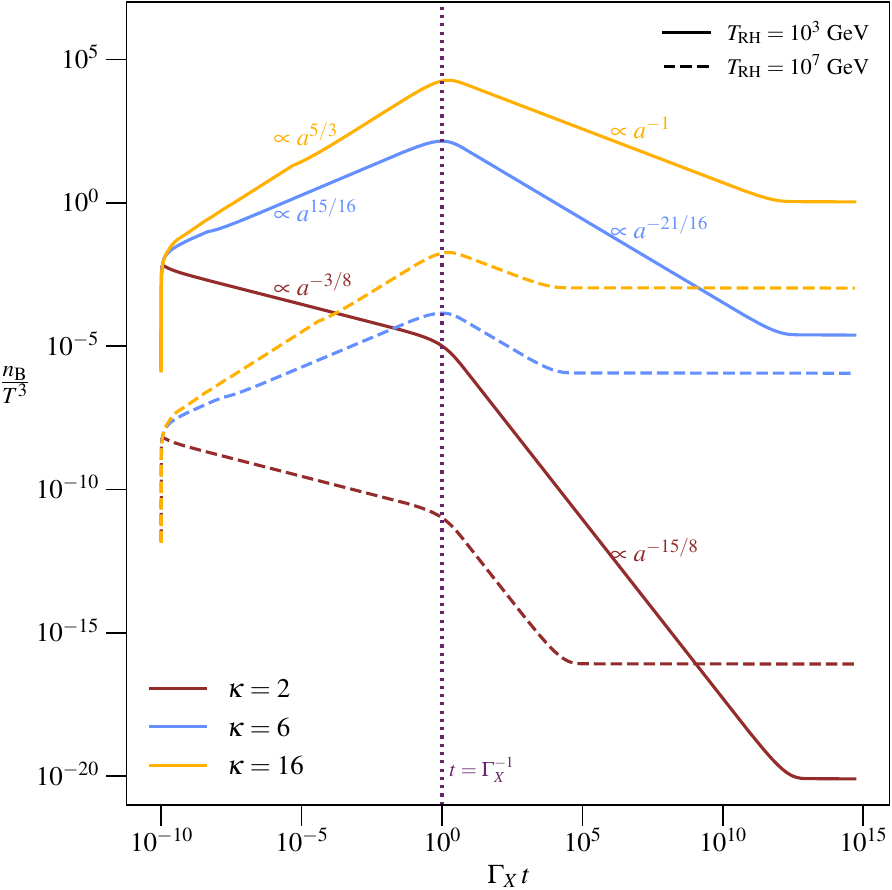}
    \caption{Ratio $n_{\rm B}/T^3$ as function of $\Gamma_X\, t$ for values of $\kappa=2$ (dark red), $\kappa=6$ (light blue) and $\kappa=16$ (light orange) as function of $\Gamma_X t$, fixing $a_{\rm in} = a_e$,  initial condition $a_{\rm in}^3 n_{X}^{\rm in} = 0.15 a_e^3 H_e^3$ for all values of $k$, $\Gamma_X t_{\rm in} = 10^{-10}$, $\epsilon_{\rm CP} = 1$, for two diferent reheating temperature values: $10^3~\GeV$ (full lines), $10^7~\GeV$ (dashed). We assumed here the decay of an out-of-equilibrium particle $X$ as source of the baryon asymmetry during reheating of an $\alpha-$attractor T-model with potential given in Eq.~\eqref{eq:potential_inflaton}. }
\label{fig:ratio}
\end{figure}

Before we delve into the specifics of how a matter-antimatter asymmetry is generated within the CGPP framework, it is important to explore the general evolution of a baryon number density produced during reheating within $\alpha-$attractor T-models of inflation. 
This will help us understand the main features of our scenario once we determine the nature of the decaying particle generated via CGPP.
Neglecting for the moment any other source of asymmetry and possible washouts, and assuming late reheating, we track the dynamics of the $n_{\rm B}$ asymmetry number density produced from the out-of-equilibrium decay of a non-relativistic particle $X$ using the following equations
\begin{align}\label{eq:simple_baryo_BEqs}
        \frac{d (n_{X} a^3)}{dt} &= -\Gamma_X n_{X},\quad 
        \frac{d (n_{\rm B} a^3)}{dt} = \epsilon_{\rm CP} \Gamma_X n_{X},
\end{align}
being $\Gamma_X$ the decay width of the $X$ particle and $\epsilon_{\rm CP}$ the CP violation parameter associated to the decay of $X$ into SM degrees of freedom.
Such equations are solved by
\begin{subequations}
    \begin{align}
        n_{X} &= n_{X}^{\rm in} \left(\frac{a_{\rm in}}{a}\right)^3 \exp(- \Gamma_X t),\\
        n_{\rm B} &= \epsilon_{\rm CP}\, n_{X}^{\rm in}\, \left(\frac{a_{\rm in}}{a}\right)^3\, (1-  \exp(- \Gamma_X t)),
    \end{align}
\end{subequations}
with $n_{X}^{\rm in}$ the initial number density for the $X$ particle at some initial scale factor $a_{\rm in}$.
At early times, $t \ll \Gamma_X^{-1}$, and depending on the equation-of-state of the Universe $w_\varphi$ during reheating, the redshift of $n_{\rm B}$ can vary significantly compared to the SM bath since
\begin{align}
    n_{\rm B} &\approx \epsilon_{\rm CP}\, n_{X}^{\rm in}\, \left(\frac{a_{\rm in}}{a}\right)^3 \left(\Gamma_X t\right)\notag\\
    &\approx \epsilon_{\rm CP}\, n_{X}^{\rm in}\, \Gamma_X\, \left(\frac{a_{\rm in}}{a}\right)^{6/(2+\kappa)},
\end{align}
where we used that $a\propto t^{2/3(1+w_\varphi)}$, and $w_\varphi=(\kappa-2)/(\kappa+2)$ as discussed before.
Utilizing the proportion of $n_{\rm B}$ to $T^3$ as a proxy for the observed baryon-to-photon ratio, and considering that $T$ scales as $a^{-3\kappa/(4(2+\kappa))}$\footnote{Note that this dependence is obtained assuming the inflaton width to be constant.} during reheating~\cite{Bernal:2019mhf},
we have
\begin{align}
    \frac{n_{\rm B-L}}{T^3} \propto a^{3 (3\kappa - 8)/(4(2+\kappa))}, \qquad \text{for } t\ll \Gamma_X^{-1}.
\end{align}
We illustrate this ratio in Fig.~\ref{fig:ratio} for values of $\kappa=2$ (dark red), $\kappa=6$ (light blue) and $\kappa=16$ (light orange) as function of $\Gamma_X t$, fixing $a_{\rm in} = a_e$, initial condition $a_{\rm in}^3 n_{X}^{\rm in} = 0.15 a_e^3 H_e^3$ for all values of $\kappa$, $\Gamma_X t_{\rm in} = 10^{-10}$, and $\epsilon_{\rm CP} = 1$, and for two values of the reheating temperature of $T_{\rm RH} = 10^7~\GeV$ (full) and $10^3~\GeV$ (dashed).
For a fixed $T_{\rm RH}$, we observe that at early times the $B$ number density experiences a more rapid redshift compared to photons for $\kappa \leq 8/3$, resulting in a suppression of the final asymmetry. Conversely, when $\kappa \geq 8/3$, $n_{\rm B}$ undergoes a slower redshift than photons. Consequently, the ratio $n_{\rm B-L}/T^3$ increases as a function of the scale factor $a$. 
Such growth ceases around $t \sim \Gamma_X^{-1}$, where the exponential term begins to diminish, causing $n_{\rm B}$ to redshift as matter, $n_{B-L} \propto a^{-3}$, while the photon density continues to increase as $n_\gamma\propto a ^{-9\kappa/(4(2+\kappa))}$ due to ongoing inflaton decay, leading to an entropy dilution of the pre-existing asymmetry. This implies that the ratio decreases as
\begin{align}
    \frac{n_{\rm B-L}}{T^3} \propto a^{-3(8+\kappa)/(4(2+\kappa))}, \qquad \text{for } t\gtrsim \Gamma_X^{-1}.
\end{align}
Post-reheating, both $n_{B-L} \propto a^{-3}$ and $n_\gamma\propto a^{-3}$, resulting in a constant ratio, thereby leading the baryon asymmetry, modulo a possible sphaleron factor, to converge toward its final value.
We can thus conclude that when $\kappa\leq 8/3$, generating the observed baryon asymmetry requires a substantial initial number density $n_X^{\rm in}$ to counterbalance the dual effects of the Universe's rapid expansion and entropy dilution. On the other hand, when $\kappa$ exceeds such a threshold, the entropy dilution at $t \gtrsim \Gamma_X$ can be compensated by the slower expansion of the Universe when $w>0$. Consequently, the initial number density of $X$ can be diminished, while still yielding the observed asymmetry.

When considering a different reheating temperature, two distinct effects come into play.
Firstly, for larger $T_{\rm RH}$, the initial $n_{\rm B}/T^3$ is smaller. This occurs because the maximum plasma temperature is $\propto T_{\rm RH}^{1/2}$~\cite{Giudice:2000ex}. Consequently, if the asymmetry originates from a non-thermal source, the initial asymmetry, $n_{\rm B}/T^3$ is initially larger for smaller reheating temperatures.
However, if asymmetry production concludes before the end of reheating, the entropy dilution resulting from inflaton decay is much more pronounced for lower $T_{\rm RH}$, leading to an additional reduction of the asymmetry.
To comprehend the dependence on $\kappa$, let us explicitly estimate the photon-to-baryon ratio calculated at reheating,
\begin{align}
    \eta_B &\equiv \frac{a^3 n_B}{a^3 n_\gamma},\notag\\
    &= \frac{\pi^2}{2 \zeta(3)} \left(\frac{\pi^2 g_{\star}}{90 M_{\rm Pl}^2}\right)^{\frac{2+\kappa}{2\kappa}} H_e^{\frac{2(\kappa-1)}{2\kappa}} T_{\rm RH}^{\frac{4-\kappa}{\kappa}} \left(\frac{a^3 n_B}{a_e^3 H_e^3}\right),
\end{align}
Therefore, we note that the interplay of all aforementioned effects results in a decrease in asymmetry for $\kappa<4$ with smaller $T_{\rm RH}$, while the opposite occurs when $\kappa>4$. When $\kappa=4$, we find that the combined effects lead to the same asymmetry. 

However, there are additional effects that cannot be neglected to correctly predict the final asymmetry.
For instance, inverse decays can washout part of the asymmetry depending on the plasma temperature.
Thus, in what follows, we solve more general equations than those presented in Eqs.~\eqref{eq:simple_baryo_BEqs} containing washout processes and the unavoidable production in the SM thermal bath. 
Such equations are
\begin{subequations}\label{eq:baryo_BEqs}
\begin{align}
    aH\frac{d N_X^{\rm th}}{da} &= - (N_X^{\rm th} - N_X^{\rm eq}) \Gamma_X^T,\\
    aH\frac{d N_X^{\rm CGPP}}{da} &= - \Gamma_X N_X^{\rm CGPP} ,\label{eq:baryo_BEqs_CGPP}\\
    aH\frac{d N_{\rm B}}{da} &= \epsilon_{\rm CP}\left[(N_X^{\rm th} - N_X^{\rm eq}) \Gamma_X^T + N_X^{\rm CGPP} \Gamma_X\right] \notag\\
    &\quad - c_X\Gamma_X^T  \frac{N_X^{\rm eq}}{N_f^{\rm eq}}N_{\rm B},
\end{align}
\end{subequations}
where $N_X^{\rm th}, N_X^{\rm CGPP}, N_X^{\rm eq}$ denote the comoving ($N_X \equiv a^3 n_X$) number densities of $X$ particle generated by the thermal plasma, CGPP, and equilibrium abundance, respectively, while $N_{B}$ represents the comoving number density of the $B$ asymmetry. $\Gamma_X^T$ refer to the thermally-averaged decay widths.
$c_X$ is a factor dependent on the coupling between the particle $X$ to the SM degrees of freedom, and $N_f^{\rm eq}$ is the equilibrium abundance of the particles in which $X$ decays.
It is important to note an important caveat regarding the aforementioned Boltzmann equations.
We assume that there is a separation of time scales between CGPP and decay processes, so that CGPP fully completes before a time $t \sim \Gamma_X^{-1}$.

In the scenario where the decay of the $X$ particle produces a $B-L$ asymmetry instead, sphalerons are expected to convert part of such asymmetry into a $B$ asymmetry. 
Such a conversion is made by including a sphaleron factor $a_{\rm sph}$, such that
\begin{align*}
    n_{\rm B} = a_{\rm sph} n_{\rm B-L}= \frac{28}{79} n_{\rm B-L},
\end{align*}
where $n_{\rm B-L}$ is the $B-L$ asymmetry number density and we have taken the SM sphaleron factor to be $a_{\rm sph} = 28/79$~\cite{Fong:2020fwk}.

Hence, for the generation of baryon asymmetry through CGPP, two main avenues emerge. Firstly, within the $\alpha-$attractor model with $\kappa \leq 8/3$, the decaying particle must be a spin-0 field with a non-conformal coupling $\xi \neq 1/6$, given its enhanced CGPP production. Alternatively, one could explore inflaton potentials characterized by $\kappa >8/3$, coupled with a decaying fermionic particle. The latter possibility will be investigated further in the subsequent section, particularly in the context of leptogenesis.

\subsection{GUT Baryogenesis}

\begin{figure*}[!t]
\centering
\includegraphics[width=0.85\linewidth]{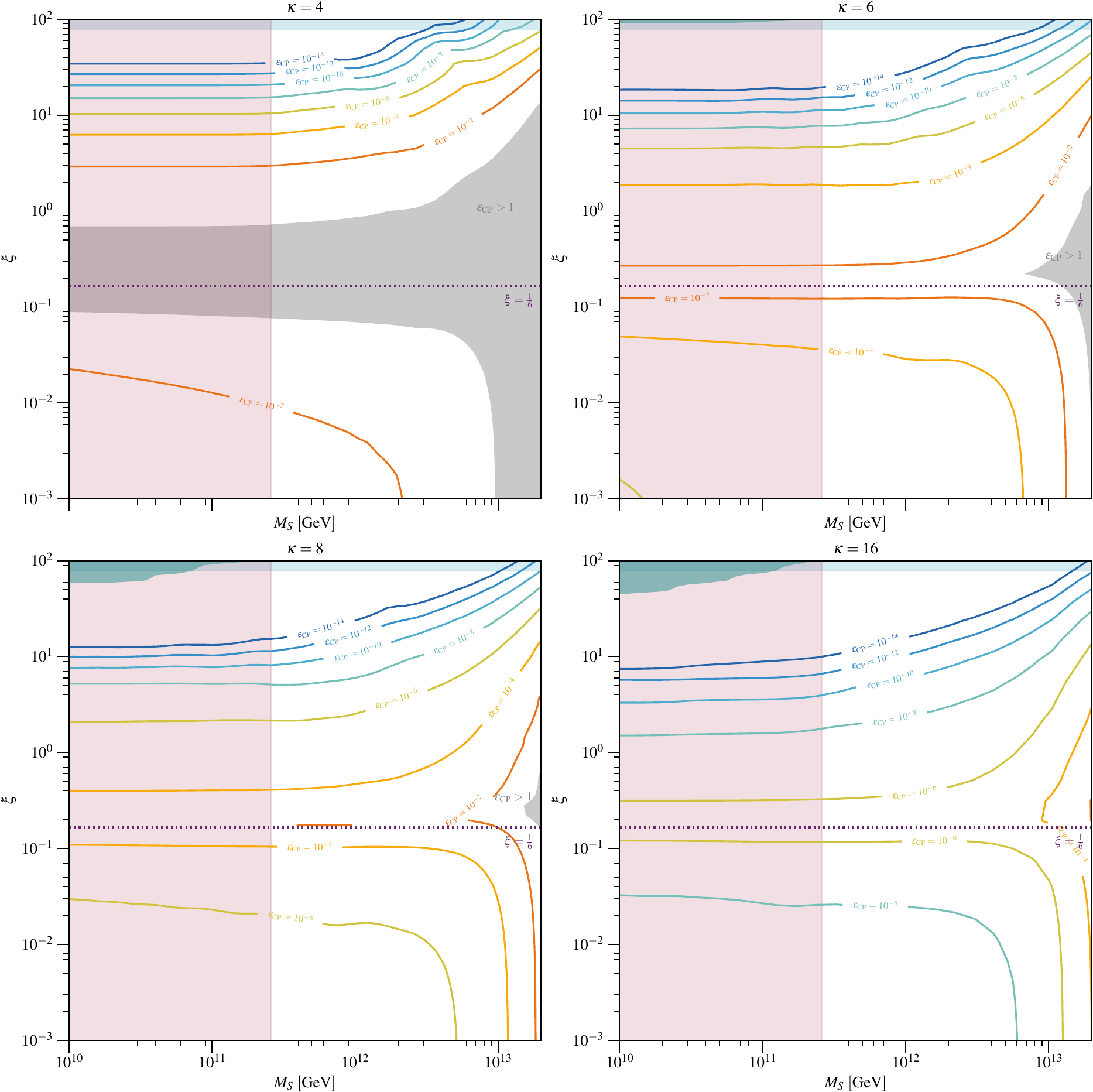}
\caption{GUT baryogenesis from scalars generated via CGPP. We present the required CP violation parameter $\epsilon_{\rm CP}$ to reproduce the observed asymmetry in the plane spanned by the scalar conformal coupling $\xi$ as function of their mass $M_S$ for different values of $\kappa =4$ (top left), $\kappa=6$ (top right), $\kappa=8$ (bottom left) and $\kappa=16$ (bottom right). The grey regions indicate where values of $\epsilon_{\rm CP} > 1$ would be needed. The light red region corresponds to values of the scalar mass that might be in conflict with proton decay. The green regions display the parameters where the scalar dominates the Universe prior decay, and the blue regions point out values where backreaction effects becomes relevant. Finally, the purple dashed line indicates the value $\xi = 1/6$.}
\label{fig:baryo}
\end{figure*}

A first possibility for the genesis of the matter-antimatter asymmetry within the framework of CGPP involves a scenario wherein the $X$ particles belong to a GUT multiplet. In this scenario, their decays would violate baryon ($B$) or baryon minus lepton number ($B-L$)\footnote{We focus here on $B-L$ violation instead of purely $B$ violation to avoid washout from sphalerons}, C, and CP conservation. Baryon asymmetry generation in GUTs suffers from several challenges, as, for instance, very massive particles $M_H\sim{\cal O}(10^{13}~\GeV)$ would have a significantly suppressed thermal production if the Universe reheated to a temperature much lower than the GUT scale. As particle generation in an expanding Universe is an irreducible production channel, CGPP offers an alternative to revive GUT baryogenesis in a minimal manner.

It is crucial to note that within GUTs, there could be a large sector of scalar bosons whose decays produce a net value of $B-L$~\cite{Rubakov:2017xzr,Hooper:2020otu}. Thus, we anticipate a viable production of baryon asymmetry across various coupling $\xi$ values.
For definiteness, we consider a simplified GUT baryogenesis model where a color-antitriplet scalar $S$ decays violating $B-L$ via $S \to qq, \bar{S} \to \bar{q}\bar{q}$~\cite{Rubakov:2017xzr}. The $S$ decay width is parametrized as
\begin{align}
    \Gamma_S = \frac{g^2}{4\pi} M_S,
\end{align}
with $M_S$ the scalar mass, and $g$ some unspecified but perturbative coupling. 
To fulfill the time separation condition between the CGPP and $S$ decay, we require the $g \lesssim 10^{-2}$ couplings to be small enough to suppress the decay width.

We solve numerically the system of equations in Eq.~\eqref{eq:baryo_BEqs} for scalar GUT baryogenesis using the infrastructure of \texttt{ULYSSES}~\cite{Granelli:2020pim,Granelli:2023vcm}.
For this we take $c_X = 4/3$ and $N_f^{\rm eq}$ to be the quark equilibrium abundance.
As an initial condition for Eq.~\eqref{eq:baryo_BEqs_CGPP}, we set $N_S^{\rm CGPP} (a_{\rm in}) = a^3 n$, where $a_{\rm in}$ corresponds to the scale factor value when $H = M_S/1000$. This choice ensures the completion of CGPP, given $a_{\rm in} \gg a_{\star}$, $a_\star$ the scale factor when $H=M_S$, before the evolution of the Boltzmann equations commences.
Subsequently, we evolve the equations until the reheating process concludes, determining the baryon asymmetry using entropy conservation.

Assuming a Universe reheating temperature of $\TRH = 10^5$ GeV, sufficiently low to mitigate the contribution from thermal processes, we depict in Fig.~\ref{fig:baryo} the degree of CP violation $\epsilon_{\rm CP}$ necessary to account for the observed matter asymmetry arising from scalars produced by CGPP, across different values of $\kappa={4,6,8,16}$ in the plane $M_S$ vs $\xi$.
The grey shaded region corresponds to nonphysical CP asymmetry values where $\epsilon_{\rm CP} >1 $, while the red region highlights values potentially excluded due to proton decay, such that $M_S \lesssim 3 \times 10^{11}~\GeV$, depending on the specific characteristics of the GUT~\cite{Hooper:2020otu}. 
For values $\xi \gtrsim 70$ (blue region), additional effects such as inflaton condensate fragmentation and backreaction on the scalar curvature may occur, as discussed in Ref.~\cite{Garcia:2023qab}, such that we would need to use more sophisticated simulations to fully determine the number density of the decaying scalars, something that lies beyond the scope of this work.
Finally, the green region in Fig.~\ref{fig:baryo} indicates the region where the scalars could dominate the evolution on the Universe before decaying.
The scenario for $\kappa=2$ is omitted, as our findings indicate that our CGPP framework could only plausibly account for the matter asymmetry at $\xi \gtrsim 40$. 

For the specific case of $\kappa=4$, we observe that scalar masses $M_S \lesssim 10^{12}~{\rm GeV} \sim 0.18 H_{e}$ require couplings $\xi \gtrsim 4$ or $\xi \lesssim 10^{-2}$, along with $\epsilon_{\rm CP} \lesssim 10^{-2}$.
The behavior for $\xi \gg 1/6$ can be elucidated by noting that for large couplings, $m_{\rm eff}^2 \approx 6 \xi (\dot{H} + 2H^2)$, resulting in a rapid mass change at the end of inflation, thereby amplifying particle production.
On the other hand, for $\xi \to 0$, $\omega^2(\eta)$ can become negative leading to a tachyonic enhancement of the particle production, see~\cite{Garcia:2023qab,Kolb:2023ydq}.
These effects lead to a large final CGPP number density that will produce the observed asymmetry for reasonable values of $\epsilon_{\rm CP}$.
For larger values of $\kappa$, there is the additional enhancement of the baryonic yield due to the reduced redshift of the $B-L$ asymmetry at early times that counterbalances the entropy dilution occurring after the scalar fully decayed, as explained in detail before. 
As consequence, we would only require values of $\epsilon_{\rm CP} \lesssim 10^{-1}$, in such a way that for $\kappa \geq 8$ almost all the parameter space becomes viable for GUT baryogenesis, including the case where the scalar couples conformally to gravity.

\begin{figure*}[!ht]
\centering
\includegraphics[width=0.925\linewidth]{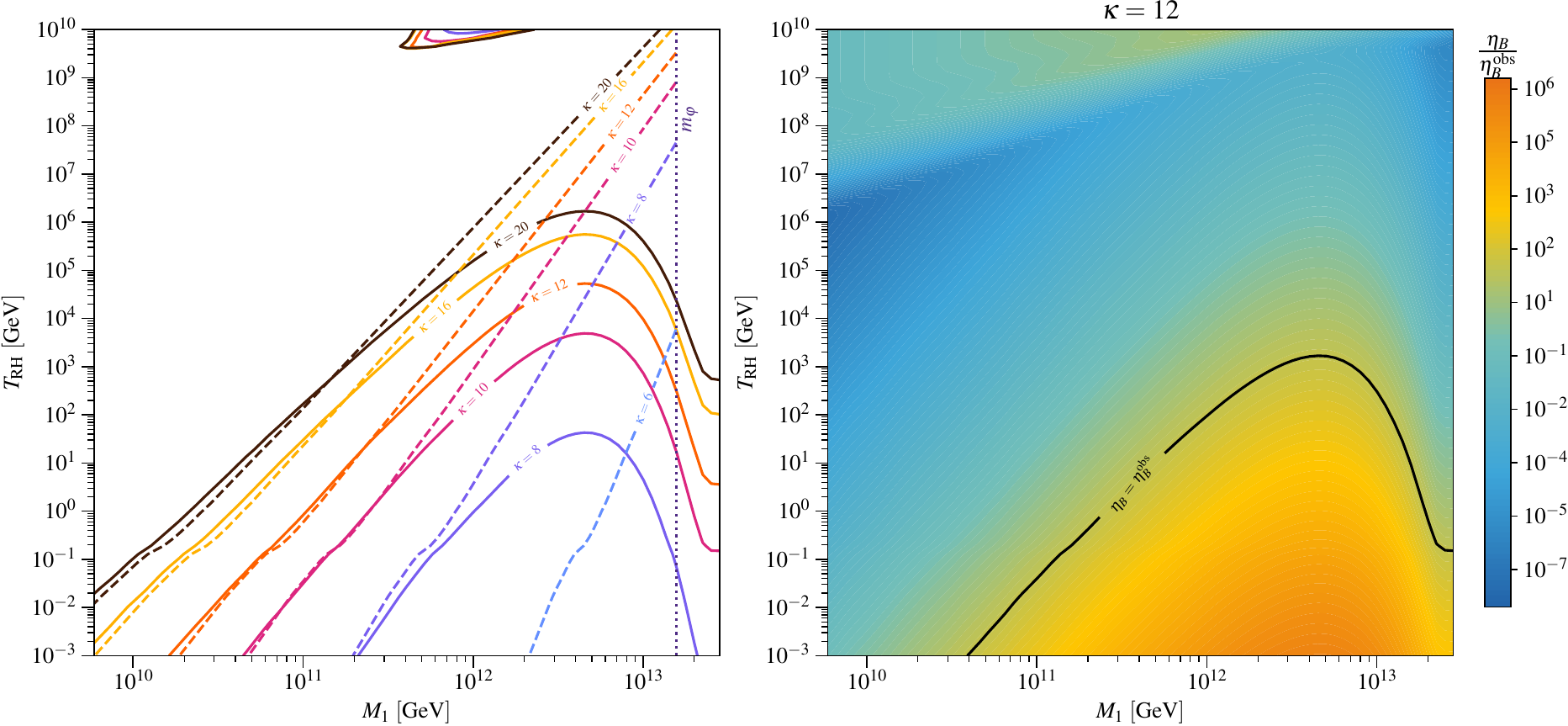}
\caption{Left -- Reheating temperature $T_{\rm RH}$ as function the right-handed neutrino mass required to produce the observed baryon-to-photon ratio for different values of $\kappa$. We compare with the results from the perturbative approach, depicted as dashed lines. Right --  Ratio of baryon-to-photon to the observed for $\kappa=12$ in the same plane $T_{\rm RH}$ vs $M_N$.}
\label{fig:lepto_CGPP}
\end{figure*}
\subsection{Leptogenesis}

Leptogenesis is one of the most appealing frameworks for explaining the surplus of matter in the early Universe, particularly due to its connection with neutrino masses~\cite{Fukugita:1986hr, DiBari:2008mp, Fong:2014gea}. In its vanilla form, heavy right-handed neutrinos (RHNs) generated thermally decay asymmetrically into leptons and Higgs particles, leading to the observed asymmetry.
To simultaneously account for neutrino masses and achieve the required baryon-to-photon ratio, the masses of the RHNs typically need to be above ${\cal O} (10^{10}~\GeV)$, assuming reasonable values for the Yukawa couplings $Y_\nu$ between the Higgs, lepton doublets, and the RHN, i.e., $Y_\nu \sim 1$. Consequently, if the Universe reheats to a temperature significantly lower than the scale of RHN masses, leptogenesis becomes unattainable.
In this context, CGPP offers a simple and feasible approach for RHN production in a cooler Universe.

To reproduce the observed neutrino mass pattern, where two non-zero neutrino quadratic mass splittings have been measured, it is necessary to introduce a minimum of two right-handed neutrinos $N_i$ ($i=1,2,\cdots$) that couple to the SM leptons and Higgs doublets through a Yukawa interaction,
\begin{align}\label{eq:lagr_ss}
    {\cal L} \supset -\overline{\ell_\alpha} H^* Y_{\alpha i} N_i - \frac{1}{2} M_{i} \overline {N_i^c} N_i +  {\rm h.c.} \,,
\end{align}
where $Y_{\alpha i}$ complex Yukawa couplings. The second term in the Lagrangian describes Majorana masses $M_i$ for the RHNs, allowed by the SM gauge group. 
In what follows, we consider the case where the RHN masses are hierarchical, $M_1 \ll M_2 \ll  M_3$, such that leptogenesis is mainly due to the out-of-equilibrium decay of $N_1$.
From this Lagrangian, we can calculate the decay width for the RHN $N_1$, denoted as $\Gamma_{1}$, which is relevant to our discussion on the generation of the baryon asymmetry. Such decay width is given by,
\begin{align}\label{eq:N1decay}
    \Gamma_{1} = \frac{1}{8\pi}(Y^\dagger Y)_{11}\, M_{1}\,.
\end{align}
Differently from the GUT baryogenesis scenario presented above, the CP violation parameter $\epsilon_{\rm CP}$ has an upper bound due to the connection of the seesaw Lagrangian in Eq.~\eqref{eq:lagr_ss} to neutrino mass parameters~\cite{Davidson:2002qv}.
Such a constraint, known as the Davidson-Ibarra bound, $|\epsilon_{\rm CP}|$ is~\cite{Davidson:2002qv}
\begin{align}
|\epsilon_{\rm CP}| \leq \epsilon_{\rm CP}^{\rm DI} \equiv \frac{3M_{1}}{16\pi v^2} \frac{\Delta m_{\rm 31}^2}{m_3 + m_1}\,,
\label{eq:DI}
\end{align}
with $\Delta m_{\rm 31}^2$ the atmospheric quadratic mass splitting measured in neutrino oscillations and $m_3\,(m_1)$ the heaviest (lightest) light neutrino mass, respectively.
For simplicity, we have assumed the Normal Ordering of neutrino masses. The results for the Inverted Ordering are anticipated to be comparable to the ones presented below.
The precise value of the CP violation parameter can be determined by employing the Casas-Ibarra parametrization~\cite{Casas:2001sr}, under the assumption that $\Delta m_{\rm 21}^2 \ll \Delta m_{\rm 31}^2$. 
These approximations imply the presence of only one complex angle, denoted as $z \equiv x + i y$, in the $R$-matrix~\cite{Hambye:2003rt}. 
As a result, the CP parameter is given by
\begin{align}\label{eq:eCP}
|\epsilon_{\rm CP}| 
= \frac{3M_{1}}{16\pi v^2}\, \frac{\Delta m_{\rm 31}^2}{m_3 + m_1}\,
\frac{\left|\sin(2x)\sinh(2y)\right|}
{\cosh(2y)-f\cos(2x)}\,,
\end{align}
being $f \equiv (m_3 - m_1)/(m_3 + m_1)$. It is clear then that the Davidson-Ibarra bound is saturated when
$x = \pm\pi/4$ and $y \to \pm\infty$. 
In thermal leptogenesis, however, a sizable $y$ results in significant washout due to inverse decays of $N_1$. Therefore, to mitigate this substantial washout while simultaneously maximizing the level of CP violation, leptogenesis would ideally take place when the washout processes from inverse decays become negligible at $T \ll M_{1}$.
Next, we concentrate on this particular situation.

Let us now examine leptogenesis stemming from CGPP within the framework of $\alpha-$attractor T-models. Similar to the methodology employed for GUT baryogenesis, we track the evolution of comoving number densities for RHNs produced thermally and via CGPP, alongside the $B-L$ asymmetry. This involves solving the system of equations in Eq.~\eqref{eq:baryo_BEqs}, setting $c_X = c_N = 1/2$, and $N_f^{\rm eq}$ representing the equilibrium number density for leptons, within the framework of \texttt{ULYSSES}~\cite{Granelli:2020pim,Granelli:2023vcm}.
Since the reheating temperature might be lower than the electroweak phase transition, we stop the generation of the asymmetry when the SM plasma has a temperature lower than $T_{\rm sp} = 131.7~\GeV$, the sphaleron freeze-out temperature~\cite{DOnofrio:2014rug}.

In contrast to the preceding subsection, we set the CP violation parameter to saturate the Davidson-Ibarra bound, i.e., $|\epsilon_{\rm CP}| = \epsilon_{\rm CP}^{\rm DI}$, and vary the reheating temperature. Our results are presented in Fig.~\ref{fig:lepto_CGPP}. In the left panel, we depict the reheating temperature as a function of the RHN mass required to reproduce the observed baryon-to-photon ratio for different values of $\kappa={6,8,10,12,16,20}$, while the dashed lines correspond to results obtained using the perturbative approach of Ref.~\cite{Co:2022bgh}. The right panel displays contours for $\eta_B/\eta_B^{\rm obs}$ for the case of $\kappa =12$.
We observe that to reproduce the observed asymmetry, lower reheating temperatures are required for $M_1 \lesssim H_{e}$. This dependence arises from the previously mentioned behavior of the asymmetry: since $n_\gamma$ is at most $\propto T_{\rm max}^3 \propto T_{\rm RH}^{3/2}$, smaller reheating temperatures result in a reduced photon number density. In contrast, the lepton asymmetry from CGPP remains independent of the plasma, causing the relative ratio to $n_\gamma$ to increase for smaller $T_{\rm RH}$, thereby enhancing the asymmetry. Consequently, to achieve the observed baryon-to-photon ratio for lighter RHNs, smaller $T_{\rm RH}$ are needed.
Moreover, for larger $\kappa$, the additional enhancement arises from the distinct redshift dependence of the lepton abundance at early times, as previously mentioned.

This behavior undergoes a change when $M_1 \sim H_{e}$. CGPP experiences an exponential suppression for $M_1 \gtrsim H_{e}$, leading to a further decrease in the required reheating temperature. 
Consistent with CGPP expectations, the highest production occurs when $M_1 \sim H_{e}$~\cite{Kolb:2023ydq}. The allowed values appearing at $T_{\rm RH} \sim 10^{10}\GeV$ for $M_1 \sim 10^{12}~\GeV$ are solely attributed to thermal leptogenesis.
These effects are clearly seen on the right panel, where we observe that the asymmetry increases for smaller $T_{\rm RH}$, while for $T_{\rm RH} \gtrsim 10^{7}~\GeV$, the effects of thermal leptogenesis become apparent.

Finally, we compare our findings with the results obtained from the perturbative approach outlined in Ref.~\cite{Co:2022bgh}. We observe that for RHN masses $M_1 \lesssim 10^{12}~\GeV$, both the perturbative and CGPP approaches exhibit similar behavior. 
However, for larger values, the perturbative approach begins to deviate from the non-perturbative CGPP results.
This is due to the fact that the perturbative approach predicts RHN number densities that scale proportionally to $T_{\rm RH}^{2(2+\kappa)/\kappa}$, extending up to the kinematic limit of $M_1 > m_\varphi$. 
As a consequence of these differences, the case $\kappa=6$, which would yield the observed asymmetry according to the perturbative approach, does not yield viable values consistent with observation in the CGPP approach.

\section{Discussion}\label{sec:Discussion}

As a consequence of the Universe's expansion post-Big Bang, the vacuum expectation value of any field becomes time-varying. This means that observers at distinct moments may detect what was previously a vacuum state as an excited state. This temporal evolution forms the core of cosmological gravitational particle production, an inherent phenomenon capable of generating heavy dark matter, which may not interact with the Standard Model sector.

In this paper, we have applied this framework for particle production to investigate the genesis of the observed matter-antimatter asymmetry in our Universe.
Taking $\alpha-$attractor T-Models for the inflaton potential as a benchmark for the inflationary period, we have identified the required parameters for two well-known baryogenesis models to account for the matter asymmetry, when the thermal production is negligible.
Initially, we e\-xa\-mi\-ned the development of a baryon-number imbalance generated during reheating in $\alpha-$attractor models. We observed that for potentials $V\propto \phi^\kappa$ with $\kappa \leq 8/3$, the $B$ number density undergoes a swifter redshift relative to photons at early times, resulting in a reduction of the final asymmetry.
Meanwhile, when $\kappa > 8/3$, the redshift of the baryon-number is less pronounced compared to photons, consequently amplifying the asymmetry.
Ultimately, the interplay between the entropy introduced by inflaton decays and the initial temperature of the Standard Model plasma, along with the aforementioned dependence of baryon asymmetry on redshift, can either amplify or diminish the resulting baryon-to-photon ratio.

We found that for $\kappa > 4$, the combined effects of these factors lead to a reduction in the asymmetry with higher reheating temperatures, whereas the opposite trend is observed for $\kappa < 4$.
This general pattern is critical for determining the necessary parameters in a given baryogenesis scenario to generate the observed asymmetry solely from CGPP.
As a result, we have found that baryogenesis from CGPP encounters significant challenges in scenarios where the Universe is matter-dominated ($\kappa=2$) during the inflaton's oscillatory phase. This tension arises from the combined effects of the redshift dependence of the baryon number and the entropy dilution resulting from inflaton decays, irrespective of the reheating temperature.

In our initial scenario, we examined baryogenesis from scalars associated with a Grand Unified Theory (GUT) with a free coupling to gravity $\xi$, which decay to generate a $B-L$ asymmetry subsequently converted to a $B$ asymmetry through sphalerons. 
We determined that for GUT baryogenesis to be feasible when the reheating temperature is significantly lower than the GUT scale, values of $\kappa \geq 4$ and CP violation parameter $\epsilon_{\rm CP} < 0.1$ are necessary. Additionally, for conformally-coupled scalars with $\xi=1/6$, we found that steeper potentials ($\kappa > 6$) are required to yield a sufficient quantity of scalars for generating the observed baryon-to-photon ratio.
As $\xi$ approaches zero, we observed that a lesser degree of CP violation is required, due to the tachyonic enhancement of scalar particle production.
Conversely, with larger values of $\xi$, the rapid change of the scalar effective mass, directly proportional to $\xi$, intensifies particle production, thereby reducing the requisite CP violation.

The second scenario that we examined involves the generation of heavy right-handed neutrinos through CGPP, whose subsequent decay generates a $B-L$ asymmetry in leptogenesis models. 
Given that the level of CP violation in this scenario is correlated with neutrino mass parameters, we investigated how CGPP leptogenesis depends on the reheating temperature. 
Assuming, for simplicity, that the Yukawa parameters in the neutrino sector saturate the Davidson-Ibarra bound, we found that, for a given $\kappa$, lower reheating temperatures necessitate lighter right-handed neutrino masses to generate sufficient asymmetry, provided these masses are smaller than the Hubble scale at the end of inflation, $H_e$.
This behavior resembles what has been previously reported in the literature, where leptogenesis is examined through the perturbative approach involving inflaton scatterings mediated by gravitons that generate right-handed neutrinos. However, when the mass of the right-handed neutrino is greater than $H_e$, the CGPP for fermions becomes less effective, causing an exponential drop in particle production, requiring lower reheating temperatures. Additionally, we found that larger values of $\kappa$ would require higher reheating temperatures.

Our framework has the potential to encompass additional baryogenesis scenarios that may encounter significant Boltzmann suppression at low reheating temperatures. Furthermore, exploring alternative types of inflationary potentials that could substantially contribute to the generation of matter over antimatter would be of interest. This is left for future  work.

\begin{acknowledgments}
The authors thank A. Kusenko, K. Petraki, A. Socha, E. Kolb, A. Long, L. Heurtier, and J. Turner for useful discussions. The work of M.M.F was supported by the European Union’s Horizon 2020 research and innovation programme under grant agreement No 101002846, ERC CoG CosmoChart. The work of Y.F.P.G has been funded by the UK Science and Technology Facilities Council (STFC) under grant ST/T001011/1. 
This project has received funding/support from the European Union’s Horizon 2020 research and innovation programme under the Marie Sk\l{}odowska-Curie grant agreement No 860881-HIDDeN.
This work has made use of the Hamilton HPC Service of Durham University.
\end{acknowledgments}


\bibliography{biblio}


\end{document}